\documentstyle[sprocl,epsf,rotate]{article}

\newcommand{\eybox}[2]{\epsfysize=#1 \rotate[l]{\epsfbox{#2}}}
\newbox\rotbox

\def\be{\begin{equation}}
\def\ee{\end{equation}}
\def\bea{\begin{eqnarray}}
\def\eea{\end{eqnarray}}

\begin{document}

\title{NOT STRANGE BUT BIZARRE PHYSICS FROM THE SAMPLE EXPERIMENT}

\author{DEREK B. LEINWEBER}

\address{Department of Physics and Math.\ Physics,
University of Adelaide 5005,
Australia\\
E-mail:~dleinweb@physics.adelaide.edu.au} 


\maketitle 

\vspace{-4.0cm}
\hfill ADP-98-54/T322
\vspace{3.5cm}

\abstracts{ Since the report of the SAMPLE Collaboration
suggesting the strange-quark contribution to the nucleon, $G_M^s(0)$,
may be greater than zero, numerous models have appeared supporting
positive values for $G_M^s(0)$.  In this paper the bizarre physics
associated with $G_M^s(0) > 0$ is illustrated.  Using new lattice
QCD results, our best estimate for $G_M^s(0)$ shifts slightly from
$G_M^s(0) = -0.75 \pm 0.30\ \mu_N$, to $G_M^s(0) = -0.62 \pm 0.26\
\mu_N$.  }

\section{Introduction}

A report of the first experimental measurement of the strange-quark
contribution to the nucleon magnetic moment, $G_M^s(0)$, was recently
published by the SAMPLE collaboration \cite{SAMPLE}.  Their findings
indicate 
\footnotetext{\noindent Presented at the workshop on ``Future
Directions in Quark Nuclear Physics,'' CSSM, Adelaide, March 9--20,
1998.  This and related papers may be obtained from:

\noindent
{\sf
http://www.physics.adelaide.edu.au/theory/staff/leinweber/publications.html}
} 
\begin{equation}
G_M^s(0.1\ {\rm GeV}^2) = +0.23 \pm 0.37 \pm 0.15 \pm 0.19\ \mu_N \, ,
\end{equation}
where the uncertainties are of statistical, systematic, and
theoretical origin respectively.\footnote{The charge of the
strange-quark is not included in the definition of $G_M^s(0)$.}  While
the uncertainties of the measurement are somewhat large and certainly
include negative values, this result suggests that $G_M^s(0)$ may
actually be positive.  Since the appearance of the SAMPLE result
\cite{SAMPLE}, numerous papers have appeared suggesting $G_M^s(0) >
0$.  It is the purpose of this discussion to point out the bizarre
properties of QCD that must be true if $G_M^s(0) > 0$, and illustrate
why future experiments are more likely to find a large negative value
closer to $G_M^s(0) \sim -0.6\ \mu_N$.

The Euclidean path integral formulation of quantum field theory is the
origin of fundamental approaches to the study of quantum
chromodynamics (QCD) in the nonperturbative regime.  An examination of
the symmetries manifest in the QCD path integral for current matrix
elements reveals various relationships among the quark sector
contributions \cite{dblEqualities}.  These relationships are
sufficient to express the strange-quark contribution to the nucleon
magnetic moment, $G_M^s(0)$, in terms of the experimentally measured
baryon magnetic moments of $p$, $n$, $\Sigma^+$, $\Sigma^-$, $\Xi^0$
and $\Xi^-$, and two ratios of quark-sector contributions to magnetic
moments.  These quark-sector ratios include the $s/d$ sea-quark loop
ratio, and either the $u_p/u_{\Sigma^+}$ ratio, expressing the
difference between a valence $u$-quark contribution to the proton and
the analogous contribution when the $u$-quark resides in the $\Sigma$
baryon, or the $u_n/u_{\Xi^0}$ ratio expressing the contribution of a
valence $u$-quark to the neutron relative to that when the $u$ quark
resides in $\Xi^0$.  In simple models, the $s/d$ sea-quark loop ratio
is given by a ratio of constituent quark masses \cite{dblEqualities}
$\sim 0.65$ while the valence ratios are assumed to be 1.  

\section{Valence Versus Sea Quarks}

   Here we carefully illustrate what is meant by sea-quark-loop
contributions and valence-quark contributions.  Finally, the
expressions for $G_M^s(0)$ are presented.

   Current matrix elements of hadrons are extracted from the
three-point function, a time-ordered product of three operators.
Generally, an operator exciting the hadron of interest from the QCD
vacuum is followed by the current of interest, which in turn is
followed by an operator annihilating the hadron back to the QCD
vacuum.  

\begin{figure}[t]
\vspace{-1.5cm}
\begin{center}
\setlength{\unitlength}{1.0cm}
\setlength{\fboxsep}{0cm}
\begin{picture}(10,5)
\put(0,0){\begin{picture}(5,5)\put(0,0){\eybox{4.0cm}
{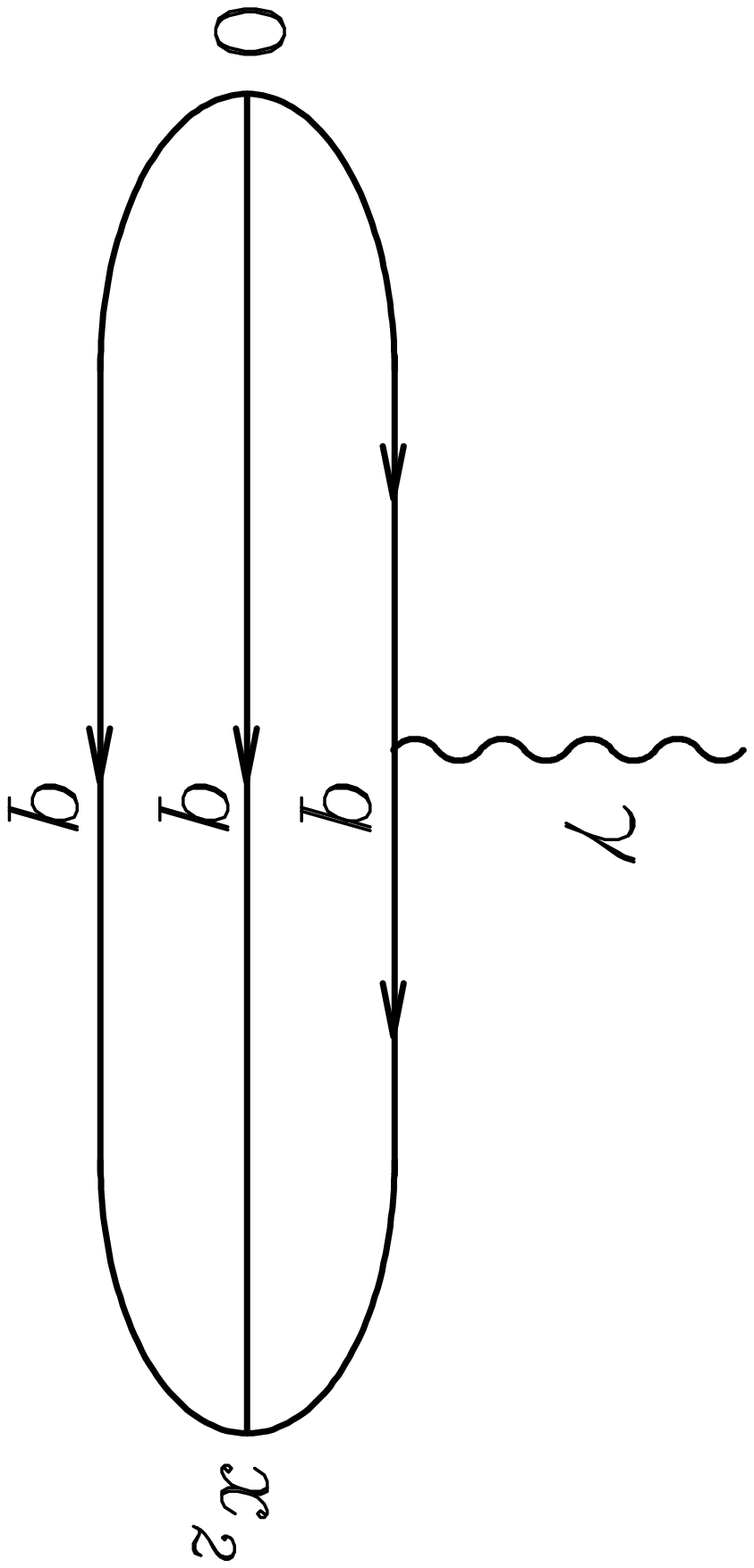}}\end{picture}}
\put(5.5,0){\begin{picture}(5,5)\put(0,0){\eybox{4.0cm}
{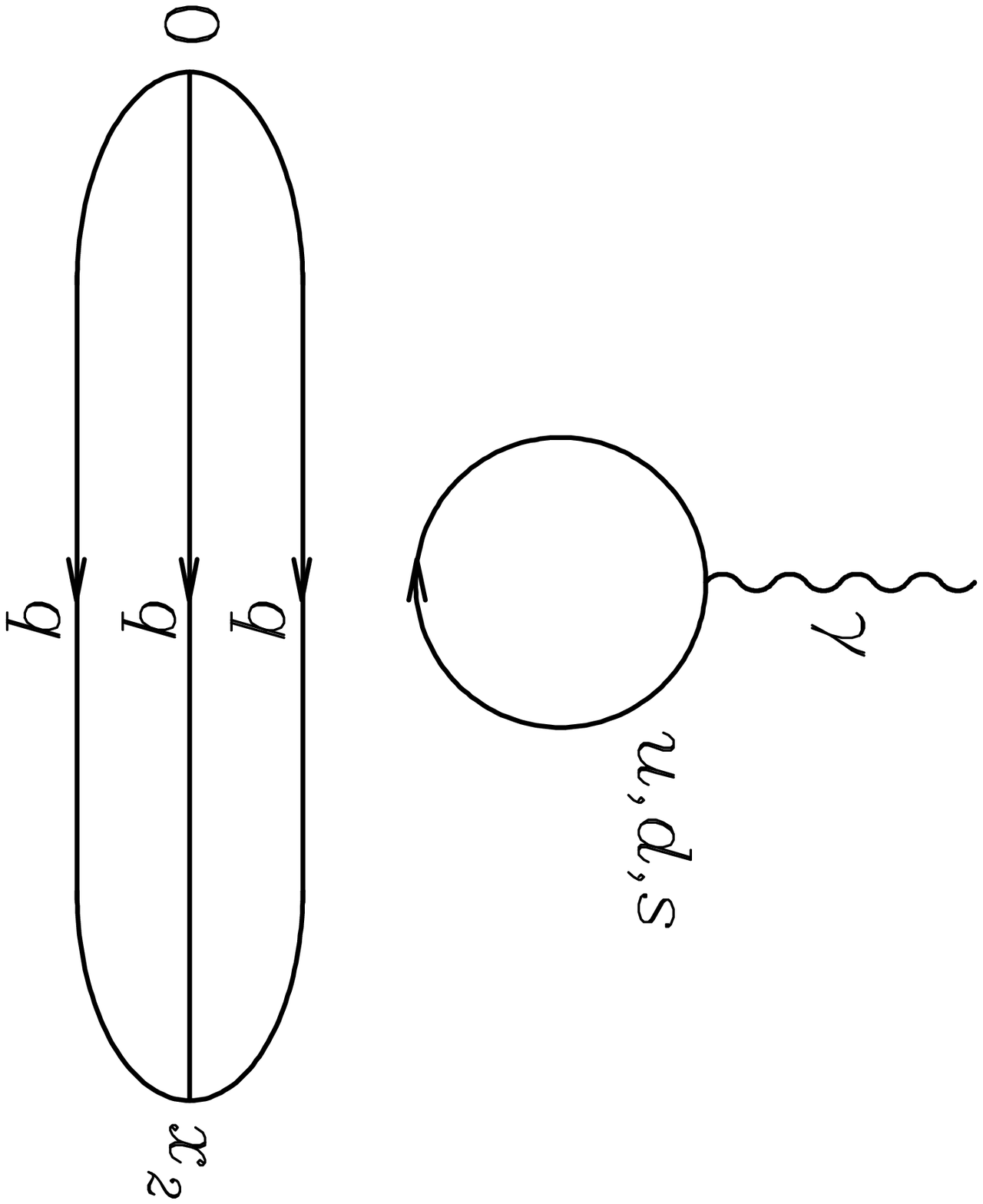}}\end{picture}}
\end{picture}
\end{center}
\caption{Diagrams illustrating the two topologically different
insertions of the current.  These skeleton diagrams for the connected
(left) and disconnected (right) current insertions may be dressed by
an arbitrary number of gluons.}
\label{topology}
\end{figure}

   In calculating the three point function, one encounters two
topologically different ways of performing the current insertion.
Figure \ref{topology} displays skeleton diagrams for these two
insertions.  These diagrams may be dressed with an arbitrary number of
gluons.  The left diagram illustrates the connected insertion of the
current to one of the valence\footnote{Note that the
term ``valence'' used here differs with that commonly used
surrounding discussions of deep-inelastic structure functions.  Here
``valence'' simply describes the quark whose quark flow line runs
continuously from $0 \to x_2$.  These lines can flow backwards as well
as forwards in time and therefore have a sea contribution associated
with them \protect\cite{dblPiCloud}.}
quarks of the baryon.  It is here that Pauli-blocking in the sea
contributions is taken into account.  The right-hand diagram accounts
for the alternative time ordering where the probing current first
produces a $q \, \overline q$ pair which in turn interacts with the
valence quarks of the baryon via gluons.

   The symmetry of the three-point correlation functions for octet
baryons having two identical quark flavors, provides the following
equalities for electromagnetic current matrix elements
\cite{dblEqualities}
\begin{eqnarray}
p =& e_u\, D_N + e_d\, S_N + O_N  \, , \qquad
n &= e_d\, D_N + e_u\, S_N + O_N  \, ,  \nonumber \\
\Sigma^+ =& e_u\, D_\Sigma + e_s\, S_\Sigma + O_\Sigma  \, , \qquad
\Sigma^- &= e_d\, D_\Sigma + e_s\, S_\Sigma + O_\Sigma  \, , \nonumber  \\
\Xi^0 =& e_s\, D_\Xi + e_u\, S_\Xi + O_\Xi  \, ,  \qquad
\Xi^- &= e_s\, D_\Xi + e_d\, S_\Xi + O_\Xi  \, .
\label{equalities}
\end{eqnarray}
Here, $D$, $S$, and $O$ represent contributions from the doubly
represented valence-quark flavor, the singly represented valence
flavor, and the sea-quark-loop sector depicted in the right-hand
drawing of Fig. \ref{topology} respectively.  The baryon label
represents the magnetic moment.  Subscripts allow for environment
sensitivity explicit in the three-point function \cite{dblEqualities}.
For example, the three-point function for $\Sigma^+$ is the same as
for the proton, but with $d \to s$.  Hence, a $u$-quark propagator in
$\Sigma^+$ is multiplied by an $s$-quark propagator, whereas in the
proton the $u$-quark propagators are multiplied by a $d$-quark
propagator.  The different mass of the neighboring quark gives rise to
an environment sensitivity in the $u$-quark contributions to
observables \cite{dblEqualities,dblOctet,dblMagMomSR,dblDecuplet,%
dblDiquarks,dblShedLight,dblEssential}.  This point sharply contrasts
the concept of an intrinsic-quark property which is independent of the
quark's environment.

Focusing now on the nucleon, we note that for magnetic properties,
$O_N$ contains sea-quark-loop contributions from primarily $u$, $d$,
and $s$ quarks.  In the SU(3)-flavor limit the charges add to zero and
the sum vanishes.  However, the heavier strange quark mass allows for
a nontrivial result.  By definition
\begin{eqnarray}
O_N &=& {2 \over 3} \,{}^{\ell}G_M^u - {1 \over 3} \,{}^{\ell}G_M^d -
{1 \over 3} \,{}^{\ell}G_M^s \, , \\
&=& {\,{}^{\ell}G_M^s \over 3} \left ( {1 - \,{}^{\ell}R_d^s \over
\,{}^{\ell}R_d^s } \right ) \, , \quad \mbox{where} \quad
{}^{\ell}R_d^s \equiv {\,{}^{\ell}G_M^s \over \,{}^{\ell}G_M^d} \, ,
\label{OGMs}
\end{eqnarray}
and the leading superscript $\ell$ reminds the reader that the
contributions are loop contributions.  Note that ${}^{\ell}G_M^u =
{}^{\ell}G_M^d$ when $m_u = m_d$.  In the simple quark model
$\,{}^{\ell}R_d^s = m_d/m_s \simeq 0.65$.  However, we will consider
$\,{}^{\ell}R_d^s$ in the range $-2$ to 2.

With no more that a little accounting, the disconnected sea-quark-loop
contributions to the nucleon, $O_N$ may be isolated from
(\ref{equalities}) in the following two favorable forms,
\begin{eqnarray}
O_N &=& {1 \over 3} \left \{ 2\, p + n - {D_N \over D_\Sigma} \left (
\Sigma^+ - \Sigma^- \right ) \right \} \, , \label{disconnN1}  \\
O_N &=& {1 \over 3} \left \{ p + 2\, n - {S_N \over S_\Xi} \left (
\Xi^0 - \Xi^- \right ) \right \} \, . \label{disconnN2}  
\end{eqnarray}
In terms of valence-quark flavors, ${D_N / D_\Sigma} = {u_p /
u_\Sigma} \equiv {d_n / d_\Sigma}$ and ${S_N / S_\Xi} = {d_p / d_\Xi}
\equiv {u_n / u_\Xi}$.  In many quark models, these ratios are simply
taken to be one.  However, we will consider the range 0 to 2.
Equating (\ref{OGMs}) with (\ref{disconnN1}) or (\ref{disconnN2})
provides
\begin{equation}
G_M^s = \left ( {\,{}^{\ell}R_d^s \over 1 - \,{}^{\ell}R_d^s } \right ) \left [
2 p + n - {u_p \over u_{\Sigma^+}} \left ( \Sigma^+ - \Sigma^- \right
) \right ] \, ,
\end{equation}
and
\begin{equation}
G_M^s = \left ( {\,{}^{\ell}R_d^s \over 1 - \,{}^{\ell}R_d^s } \right ) \left [
p + 2n - {u_n \over u_{\Xi^0}} \left ( \Xi^0 - \Xi^- \right
) \right ] \, .
\end{equation}
Incorporating the experimentally measured baryon moments indicates
\begin{equation}
G_M^s = \left ( {\,{}^{\ell}R_d^s \over 1 - \,{}^{\ell}R_d^s } \right ) \left [
3.673 - {u_p \over u_{\Sigma^+}} \left ( 3.618 \right ) \right ] \, , 
\label{ok}
\end{equation}
and
\begin{equation}
G_M^s = \left ( {\,{}^{\ell}R_d^s \over 1 - \,{}^{\ell}R_d^s } \right ) \left [
-1.033 - {u_n \over u_{\Xi^0}} \left ( -0.599 \right ) \right ] \, ,
\label{great}
\end{equation}
where the moments are in units of nuclear magnetons $(\mu_N)$.  These
expressions for $G_M^s(0)$ are perfectly valid for equal $u$ and $d$
current-quark masses and involve no other approximations.  Equation
(\ref{great}) provides a particularly favorable case for the
determination of $G_M^s$ with minimal dependence on the valence-quark
ratio.  In this short discussion, we will focus on (\ref{great})
alone.  A discussion of (\ref{ok}) will appear in a more comprehensive
publication \cite{dblInprep}.

\section{Bizarre Properties}

Here we illustrate the bizarre properties of QCD that must be true if
$G_M^s(0) > 0$.  Moreover, we'll see how a large negative
strange-quark moment the order of \cite{dblEqualities} $G_M^s = -0.75\
\mu_N$ provides results more in accord with our understanding of the
properties of QCD.

\begin{figure}[tb]
\begin{center}
\epsfxsize=10.0truecm
\leavevmode
\epsfbox{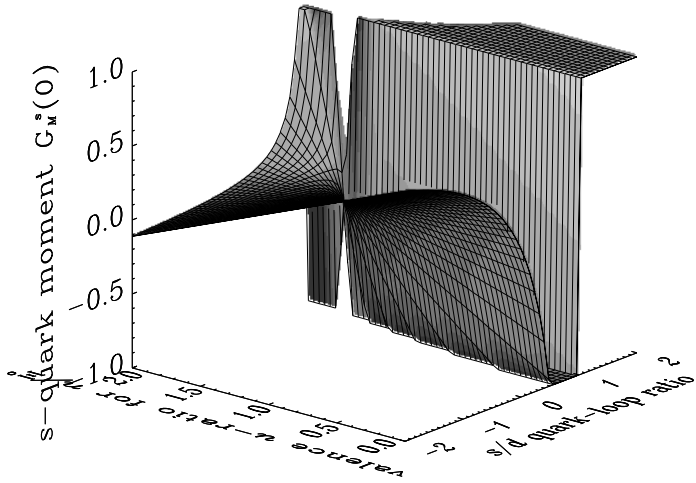}
\end{center}
\caption{The surface for $G_M^s(0)$ determined by equation
(\protect\ref{great}).  }
\label{GsMnx}
\end{figure}

Defining $G_M^s(0)$ in terms of two quark-sector ratios allows
one to plot the surface for $G_M^s(0)$ and study its properties as a
function of the quark-sector ratios.  Figure \ref{GsMnx} illustrates the
surface for $G_M^s(0)$ based on (\ref{great}).  

Since the $s$ quark is heavier than the $d$ quark, it would be very
unusual to find an enhancement in the $s$-quark-loop moment relative
to the $d$-quark-loop moment.  Such an occurrence would place the
loop-ratio ${}^{\ell}R_d^s$ greater than one.  Likewise, the mass
effect is not expected to change the sign of the $s$-quark-loop moment
relative to the $d$-quark-loop moment, but rather simply suppress the
$s$-quark-loop moment relative to the $d$-quark-loop moment.  Hence,
the region of interest along the $s/d$ quark-loop ratio axis is the
range $0 \to 1$, and it would be bizarre to find the ratio outside of
this range.

Similarly, the ratio of valence $u$-quark moments depending on whether
the $u$ quark is in an environment of two $d$ quarks or two $s$
quarks, $u_n/u_\Xi$, is expected to be the order of one.  It is
important to understand that this is a secondary effect in the total
baryon magnetic moment.  The main difference between the $n$ and
$\Xi^0$ magnetic moments is the difference between the magnetic
moment contributions of $s$ versus $d$ quarks.

\begin{figure}[t]
\begin{center}
\epsfxsize=10.0truecm
\leavevmode
\epsfbox{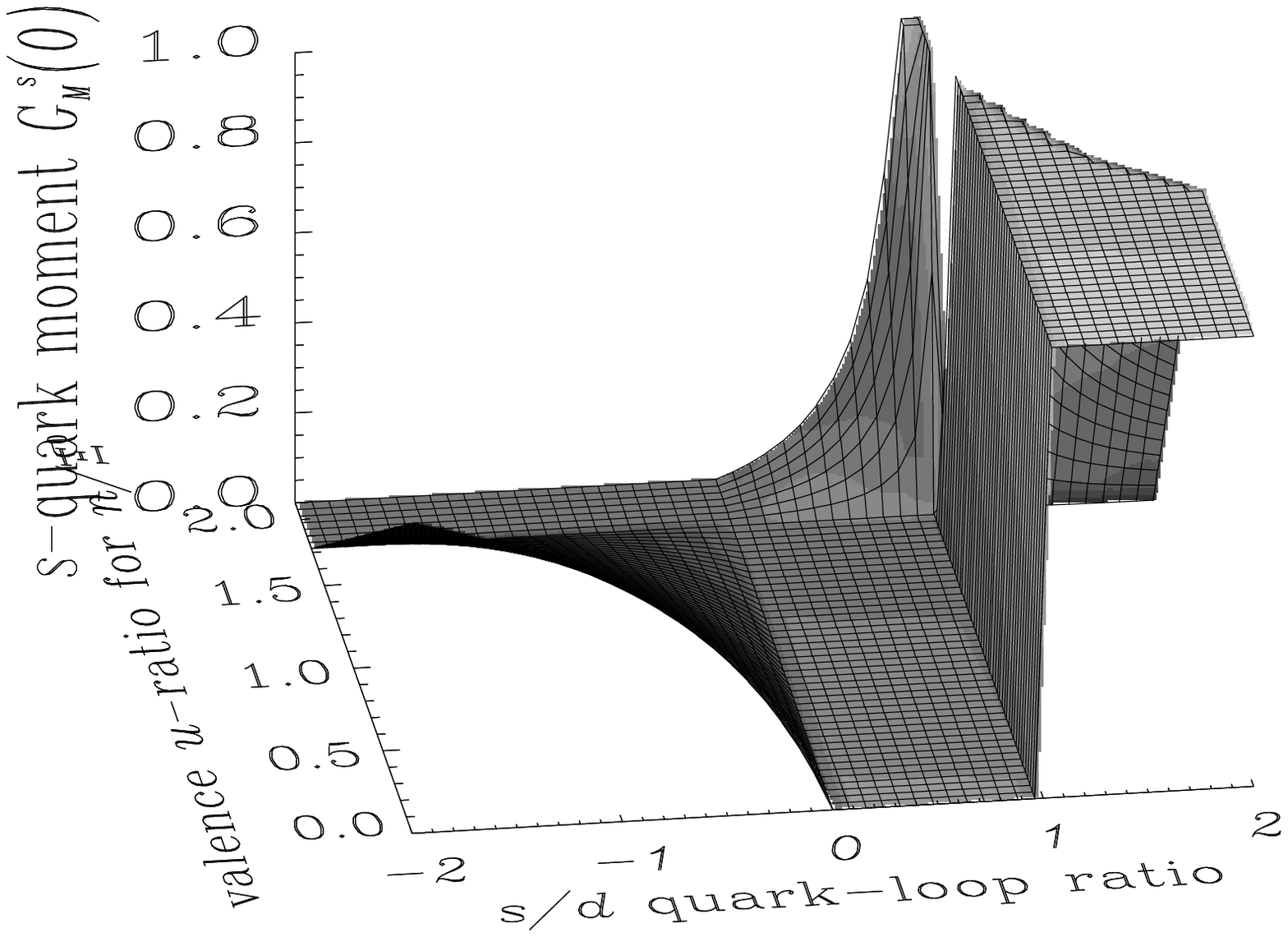}
\end{center}
\caption{The surface for $G_M^s(0)\ge 0$ determined by equation
(\protect\ref{great}).  }
\label{GsMnxPos}
\end{figure}

Figure \ref{GsMnxPos} illustrates the surface of (\ref{great}) where
$G_M^s(0) \ge 0$.  We see that the solution of (\ref{great}) for the
sea-quark-loop ratio in the region $0 \to 1$ requires rather large
values for the valence-quark moment ratio $u_N/u_\Xi > 1.7$.  To
assess whether such a ratio is reasonable, we turn to ratios of the
baryon masses and to best estimates from lattice QCD.

\begin{figure}[tb]
\begin{center}
\epsfysize=10.0truecm
\leavevmode
\rotate[l]{\epsfbox{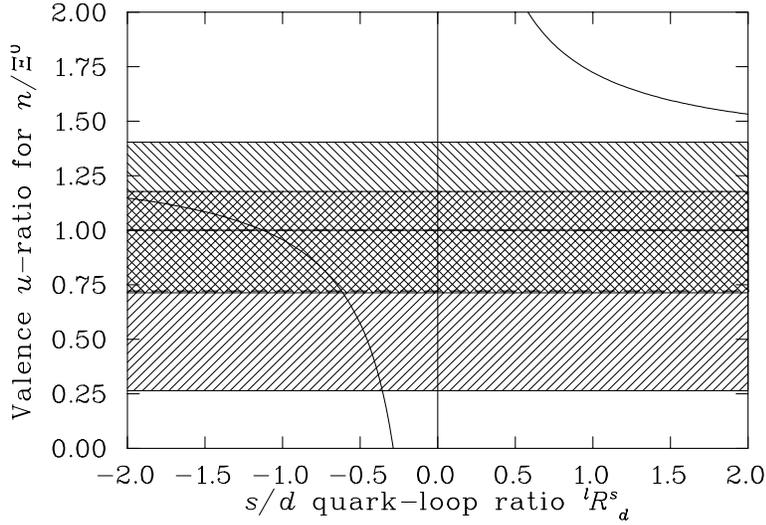}}
\end{center}
\caption{ The intersection of Figure \protect\ref{GsMnxPos} and a
horizontal plane at the central experimental value of $G_M^s(0) =
+0.23$.  Shaded regions are described in the text.}
\label{Cut23Xi}
\end{figure}

Figure \ref{Cut23Xi} displays the intersection of Figure \ref{GsMnxPos}
and a horizontal plane at the experimental central value of $G_M^s(0)
= +0.23$.  As such, it maps out the relationship between the two
quark-moment ratios which reproduces the experimental result.  This
relationship is indicated by the two curves in the upper-right and
lower-left. 

The upper region filled by backward slashes is the region between the
baryon-mass ratios $M_n/M_\Xi$ and $M_\Xi/M_n$.  These mass ratios
reflect the direct effect of the strange quark mass in much the same
way as the quark-moment ratio $d_n/s_\Xi$.  Such an effect is expected
to be much larger than the secondary effect of a change in the
environment probed by $u_n/u_\Xi$.  Still the mass-ratio region fails
to intersect with the positive $G_M^s(0)$ preferred by the SAMPLE
result when ${}^{\ell}R_d^s > 0$.

   To further explore the $u_n/u_\Xi$ ratio suggested by the SAMPLE
result, we turn to the best lattice QCD calculations of environment
sensitivity for these moments \cite{dblOctet,dblDecuplet}.  Because of
the nature of the ratios involved, the systematic uncertainties in the
lattice QCD calculations are expected to be small relative to the
statistical uncertainties.  The lattice-QCD result of $u_n/u_\Xi =
0.72 \pm 0.46$ is illustrated in Figure \ref{Cut23Xi} by the lower
forward-slash filled region.  This result suggests that the ratio is
most likely less than 1 and makes the large values required to stay in
the region of interest for the sea-quark-loop ratio look unusual.

\begin{figure}[p]
\begin{center}
\epsfxsize=10.0truecm
\leavevmode
\epsfbox{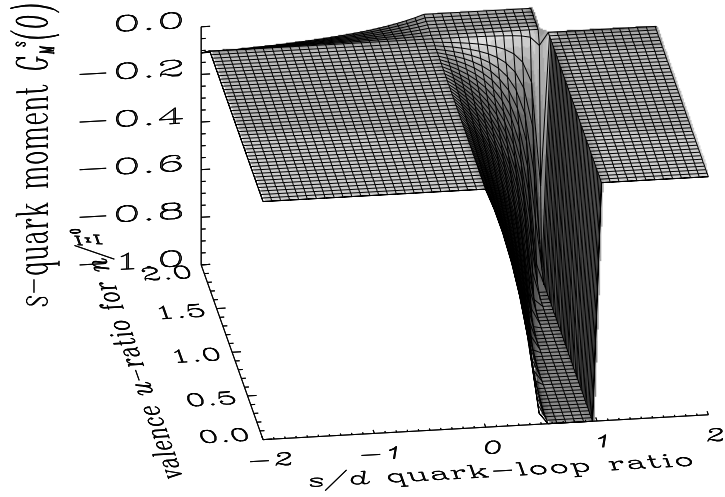}
\end{center}
\caption{The surface for $-1 \le G_M^s(0)\le 0\ \mu_N$ determined by
equation (\protect\ref{great}).  }
\label{GsMnxNeg}
\end{figure}

To contrast the bizarre physics associated with a positive value for
$G_M^s(0)$ we illustrate the surface of (\ref{great}) where $G_M^s(0)
\le 0$ in Figure \ref{GsMnxNeg}.  Here we see it is easy to satisfy the
expected properties of QCD.

\begin{figure}[p]
\begin{center}
\epsfysize=10.0truecm
\leavevmode
\rotate[l]{\epsfbox{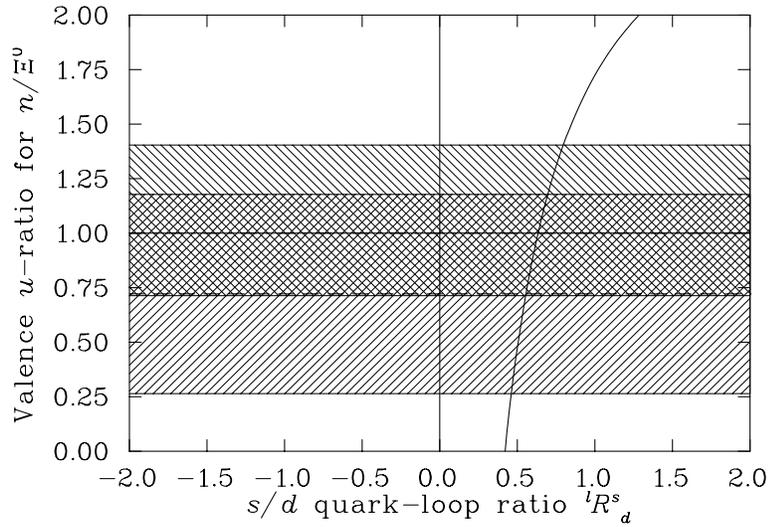}}
\end{center}
\caption{ The intersection of Figure \protect\ref{GsMnxNeg} and a
horizontal plane at the previously estimated value of
\protect\cite{dblEqualities} $G_M^s(0) = -0.75$.  Shaded regions are
described in the text.}
\label{Cut75Xi}
\end{figure}

\begin{figure}[t]
\begin{center}
\epsfxsize=10.0truecm
\leavevmode
\epsfbox{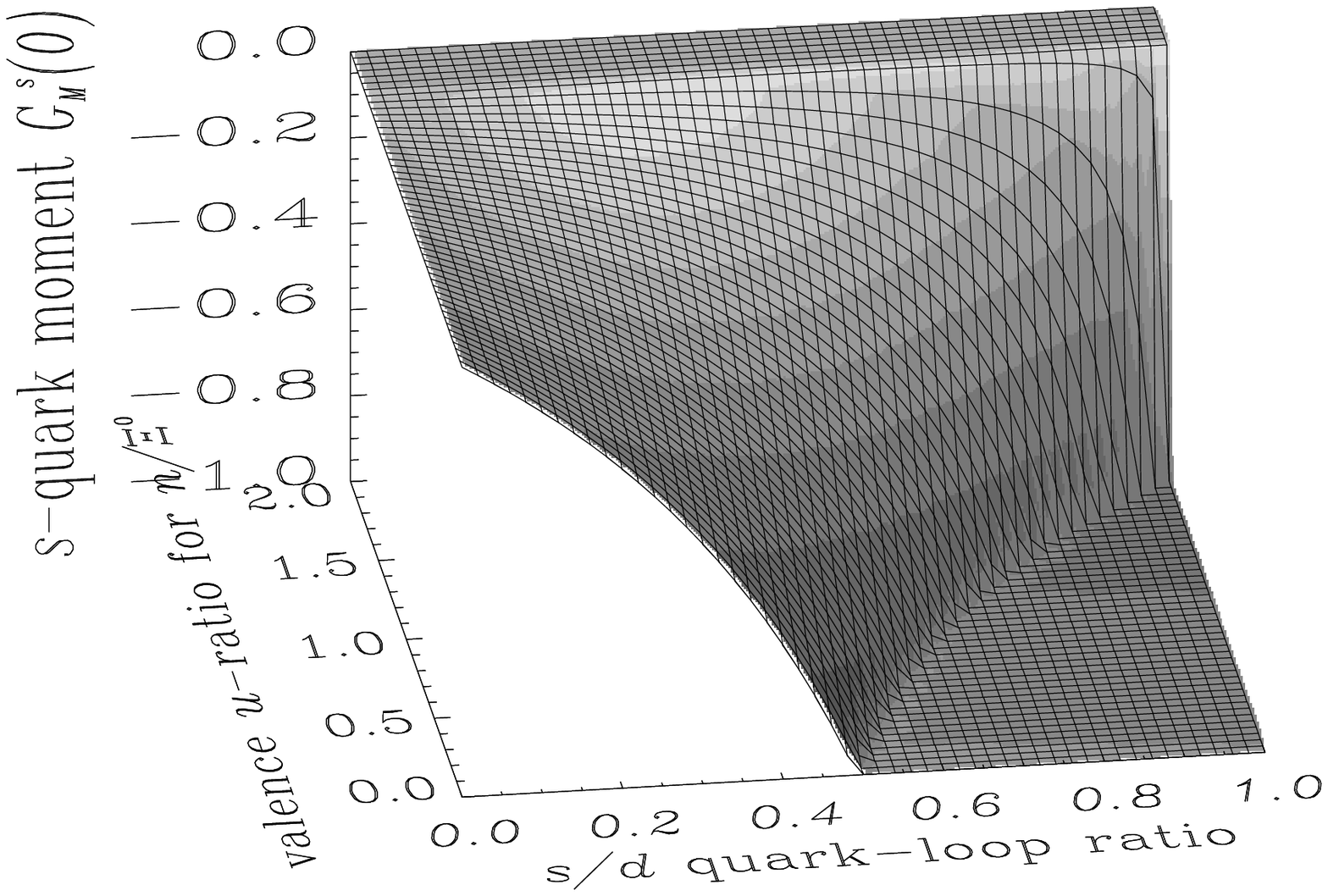}
\end{center}
\caption{ A detailed view of the surface for $-1 \le G_M^s(0) \le 0\
\mu_N$ from equation (\protect\ref{great}) in the region of interest.
}
\label{GsMnxNegLoc}
\end{figure}

Figure \ref{Cut75Xi} displays the intersection of Figure
\ref{GsMnxNeg} and a horizontal plane\break at \cite{dblEqualities}
$G_M^s(0) = -0.75$.  The curve maps out the relationship between the
two quark-moment ratios which reproduces this large negative result
for $G_M^s(0)$.  The shaded regions discussed earlier have plenty of
overlap with the regions $0 < {}^{\ell}R_d^s < 1$ and $u_n/u_\Xi \sim
1$.  Figure \ref{GsMnxNegLoc} provides a detailed view of the surface
for $-1 \le G_M^s(0) \le 0\ \mu_N$ in the region of interest.

\section{Summary Remarks and Outlook}

We have presented two equations describing the strange quark
contribution to the nucleon magnetic moment $G_M^s(0)$ in terms of the
ratio of strange to light sea-quark-loop contributions and
valence-quark ratios probing the subtle effects of environment
sensitivity.  The equations involve no approximations outside of the
usual assumption of equal current quark masses.  The sea-quark-loop
ratio probes the quark-mass suppression of the strange-quark loop
relative to the light-quark loop such that the ratio is expected to
lie between 0 and 1.  Best estimates \cite{liuDisconn} place the
ratio\footnote{Jackknife errors are not available for this ratio.  The
quoted uncertainties are based on a correlated relative-error analysis
where $y = a/b$ and $dy/y = \mid da/a - db/b \mid$.  This approach is
known to reproduce uncertainties for closely related ratios
\protect\cite{liuDisconn} where jackknife error estimates have been
performed.}
at 0.55(6).  The valence-quark ratios probing the effects of
environment sensitivity are expected to be the order of 1.  Estimates
based on SU(3) flavor symmetry breaking in the baryon masses and
lattice QCD indicate that valid solutions require $G_M^s(0)$ large and
negative. 

Positive values for $G_M^s(0)$ require the sea-quark-loop ratio to
exhibit bizarre behavior by having either the heavier quark mass
enhance the magnetic moment contribution or have the mass effect
actually change the sign of the sea-quark magnetic-moment
contribution.  Alternatively, the valence-quark ratio $u_n/u_\Xi$
probing environment sensitivity is required to be much larger than 1
and larger than conservative limits.

Using the lattice-QCD result \cite{dblOctet,dblDecuplet} of $u_n/u_\Xi
= 0.72 \pm 0.46$ and the new direct lattice-QCD estimation
\cite{liuDisconn} for the sea-quark-loop ratio ${}^{\ell}R_d^s = 0.55
\pm 0.06$, our best estimate shifts slightly from \cite{dblEqualities}
$G_M^s(0) = -0.75 \pm 0.30\ \mu_N$, to
\begin{equation}
G_M^s(0) = -0.62 \pm 0.26\ \mu_N \, .
\end{equation}

\section*{Acknowledgements}
I thank Tony Thomas and Tony Williams for insightful conversations.
Support from the Australian Research Council is also gratefully
acknowledged.

\section*{References}


\end{document}